\documentclass[
               showpacs,
               preprintnumbers,
               aps,
               prd,
               a4paper,
               superscriptaddress,
               nofootinbib,
               tightenlines,
               floats,floatfix,11pt
               ]{revtex4-2}

\usepackage{amsmath}
\usepackage{graphicx}
\usepackage{xcolor}

\begin{document}

\title{Stability Analysis of Axion Dark Energy Model}

\author{Ankan Mukherjee}
\email{ankan.ju@gmail.com}
\affiliation{Department of Physics, Bangabasi College, Kolkata - 700009, India.}

\author{Nandan Roy}
\email{nandan.roy@mahidol.ac.th (Corresponding Author)}
\affiliation{NAS, Centre for Theoretical Physics \& Natural Philosophy, Mahidol University,
Nakhonsawan Campus, Phayuha Khiri, Nakhonsawan 60130, Thailand}

\pagestyle{myheadings}
\newcommand{\be}{\begin{equation}}
\newcommand{\ee}{\end{equation}}

\begin{abstract}
We investigate the stability of an axion dark energy model characterized by a periodic scalar-field potential motivated by pseudo–Nambu–Goldstone bosons. While such models are well suited to drive late-time cosmic acceleration and can exhibit rich dynamical features, including oscillatory behavior of the dark-energy equation-of-state parameter, their viability crucially depends on the existence of stable cosmological solutions. By reformulating the background evolution equations as an autonomous dynamical system, we identify the fixed points corresponding to radiation-dominated, matter-dominated, and dark-energy–dominated epochs. Using both analytical and numerical techniques, we analyze the stability properties of these fixed points within the observationally allowed region of parameter space. Our results show that the radiation- and matter-dominated phases emerge naturally as saddle points, ensuring a standard cosmological evolution, while the late-time dark-energy–dominated solution acts as a stable attractor. This guarantees the robustness of cosmic acceleration against variations in initial conditions.  Overall, our study establishes the axion dark energy model as a dynamically consistent and observationally viable candidate for explaining the late-time accelerated expansion of the Universe.

\end{abstract}

\maketitle


\section{Introduction}

The discovery of the late-time accelerated expansion of the Universe \cite{SupernovaSearchTeam:1998fmf,SupernovaCosmologyProject:1998vns} has established the need for physics beyond a matter-only cosmological model. Within General Relativity (GR), a Universe filled only with ordinary matter undergoes decelerated expansion. The observed acceleration therefore requires either a modification of GR on cosmological scales or the presence of an additional component with sufficiently negative pressure, commonly referred to as dark energy. Despite extensive theoretical and observational efforts, the physical origin and nature of dark energy remain unresolved. The simplest candidate is the cosmological constant, associated with vacuum energy density \cite{Padmanabhan:2002ji}. Its theoretical interpretation, however, is affected by the well-known fine-tuning problem. In recent years, increasing attention has also been drawn to tensions within the standard $\Lambda$CDM framework. One of the most significant is the discrepancy between early- and late-time determinations of the Hubble constant $H_0$. Early-Universe probes such as the Cosmic Microwave Background (CMB) \cite{Planck2020}, Baryon Acoustic Oscillations (BAO) \cite{BAO2017}, and Big Bang Nucleosynthesis (BBN) \cite{BBN2021}, together with large-scale-structure surveys \cite{DES2018}, favor lower values of $H_0$. In contrast, late-Universe measurements from SH0ES \cite{Sh0ES2019} and H0LiCOW \cite{Wong:2019kwg} yield higher values, with representative determinations such as those discussed in Ref.~\cite{Riess_2022}. A related tension appears in the amplitude of matter fluctuations. Measurements of $\sigma_8$ and the derived parameter $S_8=\sigma_8\sqrt{\Omega_m/0.3}$ from low-redshift probes, including weak lensing, galaxy clustering, and cluster counts \cite{Simon:2022lde,Yuan:2022jqf,DES:2021vln,Busch:2022pcx,Wright:2025xka}, tend to prefer lower values than those inferred from the CMB \cite{Planck:2015lwi}. These tensions continue to motivate extensions of the standard cosmological model, particularly in the dark-energy sector.

Recent observations from the Dark Energy Spectroscopic Instrument (DESI) further motivate the study of dynamical dark energy. Analyses based on BAO and the full shape of galaxy clustering have tested departures from a cosmological constant and possible redshift evolution of the dark-energy equation of state \cite{DESI:2024hhd,DESI:2024aqx,DESI:2024uvr,DESI:2025zgx,DESI:2025fii,Filbert:2023zhg,Ishak:2024jhs}. Within phenomenological descriptions such as the Chevallier--Polarski--Linder parametrization, combinations of DESI with CMB and supernova data allow departures from $w=-1$ \cite{Ishak:2024jhs}. At the same time, a phenomenological equation of state need not uniquely specify the underlying microphysics \cite{Cortes:2024lgw,Roy:2025cxk,Roy:2024kni,shlivko2024assessing}. This has renewed interest in physically motivated scalar-field scenarios and in understanding how observationally allowed evolution is realized dynamically \cite{Wolf:2023uno,Shlivko:2024llw,Wolf:2024eph,Thanankullaphong:2026anl}.

The cosmological behaviour of a scalar field depends crucially on its potential. Periodic dark-energy potentials have been investigated in a variety of contexts \cite{Johnson:2008se,Dutta:2008px,Joseph:2021omr,Escamilla:2024fzq,Rezaei:2024vtg}. Axion-like fields are especially well motivated because they arise as pseudo--Nambu--Goldstone bosons and naturally possess periodic potentials \cite{DuttaK:2021,Choi:2019jck}. Such fields can act as thawing dark energy: the equation-of-state parameter remains close to $-1$ when Hubble friction freezes the field and evolves away from that value when the field begins to roll. Ruchika \textit{et al.} \cite{Ruchika:2020avj} studied the late-time cosmology of an axion dark-energy potential supplemented by a constant contribution and obtained observational constraints on the parameters of the model. A complete assessment of this class of models also requires a careful analysis of their phase-space structure. The periodic potential can generate nontrivial trajectories and damped oscillations of the scalar-field equation of state. For the minimally coupled canonical scalar considered here, however, $w_\phi\geq-1$ whenever $\rho_\phi>0$, so these oscillations do not represent a genuine crossing of the phantom divide. Dynamical-systems methods provide a systematic way to identify the radiation-, matter-, and dark-energy-dominated regimes and to determine whether the late-time accelerated solution is an attractor. Dynamical-systems techniques have a long history in cosmology \cite{Wainwright:2005,Coley:2003,Roy:2026jcm}. Their applications include inflationary scalar-field models \cite{Gunzig:2000,Carot:2003}, late-time scalar-field cosmologies \cite{Urena-Lopez:2005pzi}, axion quintessence \cite{Kumar:2013oda,Sen:2009yh}, phantom, tachyon and k-essence models \cite{Fang:2014qga,Fang:2014bta}, quintessence and holographic dark energy \cite{Roy:2014yta,Roy:2014wqa}, interacting holographic dark energy \cite{Mishra:2019uqm}, and warm quintessential inflationary dark energy \cite{Das:2023rat}.

In the present work we reconsider the stability of the axion dark-energy model discussed in Ref.~\cite{Ruchika:2020avj}. We formulate the system directly in terms of the standard normalized variables $(x,y)$ and retain the periodic phase of the scalar field as an additional dynamical variable. This avoids singular coordinate transformations at $\Omega_\phi=0$ and preserves the sign of the scalar-field velocity. We also keep the exact field dependence of the logarithmic potential slope and of $\Gamma$. The resulting system allows the physical critical points and their stability to be identified analytically. In Sec.~II we describe the axion dark-energy model. In Sec.~III we formulate the autonomous system, determine its critical points and discuss their parameter dependence. Section~IV contains our conclusions.

\section{Axion dark energy}
\label{model}

The axion field is characterized by a periodic potential, which may be written in the general form
\be
V(\phi)=\lambda^4\left[1+\frac{\lambda_l^4}{\lambda^4}
\cos\left(\frac{\phi}{f_1}\right)\right],
\label{Vphi}
\ee
where $f_1$ denotes the axion decay constant, while $\lambda$ sets the characteristic energy scale of the potential. The parameter $f_1$ determines the field-space period of the oscillatory term. A larger value of $f_1$ corresponds to a more slowly varying and locally flatter potential, whereas a smaller value of $f_1$ leads to a more rapidly varying potential with a larger curvature.

In string-inspired scenarios, several axion fields with different masses and decay constants may in general be present. The heavier axion fields typically evolve toward their respective minima at earlier stages of the cosmological evolution, so that their late-time dynamical contribution becomes negligible. Their residual vacuum-energy contributions can then be absorbed into an approximately constant background term. Consequently, the late-time dynamics can be effectively described by the lightest axion field, denoted by $\phi_l$, with a potential of the form
\be
V(\phi_l)=\lambda^4+\lambda_l^4
\cos\left(\frac{\phi_l}{f_1}\right).
\ee
Equivalently, this may be written as
\be
V(\phi_l)=\lambda^4\left[
1+\frac{\lambda_l^4}{\lambda^4}
\cos\left(\frac{\phi_l}{f_1}\right)
\right],
\label{Vphil}
\ee
which is the form adopted in the subsequent analysis. Here $\phi_l$ denotes the lightest axion field, which is assumed to dominate the late-time dynamics. The potential consists of two contributions: a constant term and a periodic oscillatory term. Depending on the relative magnitudes of $\lambda$ and $\lambda_l$, two qualitatively different regimes can arise. If $\lambda_l>\lambda$, the minimum of the potential becomes negative, while the maximum remains positive. In this case, accelerated expansion may occur when the scalar field evolves near the positive-energy maximum of the potential, corresponding to a hill-top quintessence scenario \cite{Kamionkowski:2014zda,Dutta:2008qn}. On the other hand, when $\lambda>\lambda_l$, the minimum of the potential remains positive and the model corresponds to the usual quintessence branch. In this regime, the constant contribution dominates the overall energy scale of the potential, while the periodic term acts as a modulation. The axion field can then evolve toward, and oscillate around, the positive minimum of the potential. Such oscillatory dynamics can in turn induce a damped oscillatory behaviour in the dark-energy equation-of-state parameter $w_{\rm DE}$. In the present analysis, the potential is presented in slight simple fashion as   
\be
V(\phi)=V_0[1+p\cos{(\phi/f_1)}],
\label{Vphi2}
\ee
where $V_0=\lambda^4$ and $p=\lambda^4_l/\lambda^4$. The dark energy model with this potential is confronted with recent observational data sets and constraints on cosmological parameters are obtained by Ruchika et al. \cite{Ruchika:2020avj}. Both the normal quintessence and hill top quintessence scenarios are invoked there in \cite{Ruchika:2020avj}. In the following section, we carry out a detailed dynamical systems analysis of a minimally coupled scalar field with an axion potential to understand  its qualitative behaviour.

\section{Dynamical system analysis}
\label{DSsection}

\subsection{Background equations and normalized variables}
The mathematical formalism to study cosmological dynamics starts from the Friedmann-Lemaitre-Robertson-Walker metric, given for a spatially flat universe as,

\be 
ds^2=-dt^2+a^2(t)\left[dr^2 + r^2d\Omega^2 \right],
\ee
where the time-dependent term $a(t)$, the scale factor, takes care of the evolution of the spatial separation between two coordinate points. Friedmann equations govern the dynamics,  

\be
3H^2=8\pi G\rho_{tot}
\label{friedmann1}
\ee

\be
2\dot{H}+3H^2=-8\pi G p_{tot}
\label{friedmann2}
\ee
 where the Hubble parameter $H=\dot{a}/a$, overhead dot denoting time derivative, $\rho_{tot}$ and $p_{tot}$ are respectively the total energy density and pressure-like contributions from all the components present in the energy budget of the universe. The total energy density is written as

\be 
\rho_{tot}=\rho_m + \rho_r + \rho_{de}.
\ee
where $\rho_m$ is the energy density from cold dust matter, $\rho_r$ is the radiation energy density, $\rho_{de}$ is the dark energy density. For the scalar field, the energy density and corresponding pressure like contribution are given as,

\be
\rho_{de}=\rho_{\phi}=\frac{1}{2}\dot{\phi}^2+V(\phi)~~~~~~~ p_{\phi}=\frac{1}{2}\dot{\phi}^2-V(\phi), 
\ee
and the corresponding dark energy equation of state parameter is given as,

\be
w_{\phi}=\frac{\dot{\phi}^2-2V(\phi)}{\dot{\phi}^2+2V(\phi)}.
\label{wde}
\ee

From contracted Bianchi identity conservation equation of dark energy density yield as,

\be 
\dot{\rho_{\phi}}+3H(1+w_{\phi})\rho_{\phi}=0.
\label{conservationDE}
\ee

It is always convenient to use the normalized form of energy density of different components, defined as

\be
\Omega_i=\frac{8\pi G\rho_i}{3H^2},
\ee
and consequently from equation (\ref{friedmann1}),  $\Omega_m+\Omega_r+\Omega_{\phi}=1$. The matter and radiation components satisfy $\dot\rho_m+3H\rho_m=0$ and $\dot\rho_r+4H\rho_r=0$, respectively.

To study the model’s dynamical-system behavior, it is standard to introduce a set of dimensionless variables. Here we adopt the conventional choice
\be
x\equiv\frac{\kappa\dot\phi}{\sqrt6H},\qquad
y\equiv\frac{\kappa\sqrt{V}}{\sqrt3H},\qquad
\Omega_r\equiv\frac{\kappa^2\rho_r}{3H^2}.
\label{xydef}
\ee
With these newly defined dimensionless variables, the scalar-field and matter density parameters are then given by
\be
\Omega_\phi=x^2+y^2,\qquad
\Omega_m=1-x^2-y^2-\Omega_r.
\label{Omegas}
\ee
In terms of these variables,
\be
w_\phi=\frac{x^2-y^2}{x^2+y^2},\qquad
\gamma_\phi\equiv 1+w_\phi=\frac{2x^2}{x^2+y^2},
\label{wxy}
\ee
where $\gamma_\phi$ denotes the scalar-field barotropic index (rather than the equation-of-state parameter itself). From Eq.~(\ref{wxy}) it follows that a canonical scalar with positive energy density satisfies $w_\phi\geq -1$. The effective equation of state is
\be
w_{\rm eff}=x^2-y^2+\frac{\Omega_r}{3},
\label{weff}
\ee
while
\be
\frac{\dot H}{H^2}=-\frac32\left(1+x^2-y^2+\frac{\Omega_r}{3}\right).
\label{Hdot}
\ee
For the illustrative numerical evolution shown in Fig.~\ref{wDEz}, we plot the normalized Hubble parameter $h(a)\equiv H(a)/H_0$ together with the dark-energy equation-of-state parameter $w(a)$, where $a$ is the scale factor of the universe. For this figure, we choose representative values of the model parameters $p$ and $f_1$ as well as the initial conditions, in agreement with the best-fit intervals quoted in Ref.~\cite{Ruchika:2020avj}..

\IfFileExists{ha.eps}{
\begin{figure}[tb]
\begin{center}
\includegraphics[width=0.36\textwidth]{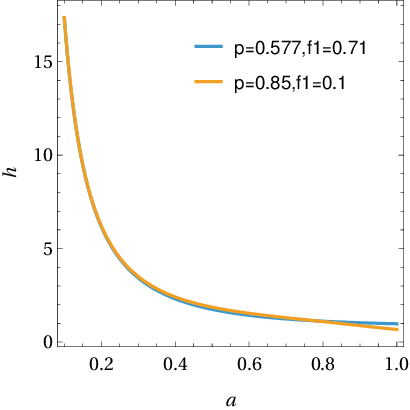}
\includegraphics[angle=0,width=0.40\textwidth]{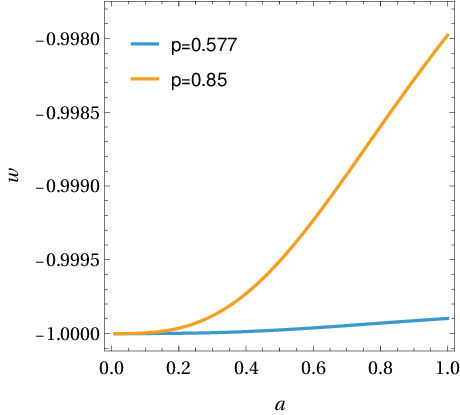}
\end{center}
\caption{{\small Left panel shows the scaled Hubble parameter $h$ as a function of scale factor $a$ for representative values of the model parameters. The right panel shows the scalar-field equation-of-state parameter $w_\phi(a)$. Other parameters are kept fixed at the best-fit values reported in Ref.~\cite{Ruchika:2020avj}.}}
\label{wDEz}
\end{figure}
}{}


For a generic scalar potential $V(\phi)$, it is useful to introduce the dimensionless logarithmic slope
\be
\Lambda_\phi\equiv-\frac{1}{\kappa V}\frac{dV}{d\phi}
\label{Lambdadef}
\ee
and the related curvature parameter
\be
\Gamma\equiv\frac{VV_{,\phi\phi}}{V_{,\phi}^2}.
\label{Gammadef}
\ee
For the axion potential in Eq.~(\ref{Vphi2}), we retain the full periodic field dependence by defining the phase
\be
\theta\equiv\frac{\phi}{f_1},\qquad F\equiv\kappa f_1.
\label{thetaF}
\ee
In terms of $\theta$ one finds
\be
\Lambda_\phi(\theta)=\frac{p}{F}\frac{\sin\theta}{1+p\cos\theta},
\label{LambdaTheta}
\ee
and
\be
\Gamma(\theta)=-\frac{\cos\theta\left(1+p\cos\theta\right)}{p\sin^2\theta}.
\label{GammaTheta}
\ee
The apparent divergence of $\Gamma$ at $\sin\theta=0$ is an artifact of its definition, which involves $V_{,\phi}^{-2}$, and does not correspond to a singularity of the underlying scalar-field equations. It is therefore preferable to evolve $\theta$ directly, rather than assume a constant $\Gamma$ or treat $\Lambda_\phi$ as an independent constant parameter.

Using $N\equiv\ln a$ as the time variable, the autonomous system becomes
\begin{align}
x'={}&-3x+\sqrt{\frac32}\,\Lambda_\phi(\theta)y^2
+\frac32x\left(1+x^2-y^2+\frac{\Omega_r}{3}\right),
\label{autox}\\
y'={}&-\sqrt{\frac32}\,\Lambda_\phi(\theta)xy
+\frac32y\left(1+x^2-y^2+\frac{\Omega_r}{3}\right),
\label{autoy}\\
\Omega_r'={}&\Omega_r\left(-1+3x^2-3y^2+\Omega_r\right),
\label{autor}\\
\theta'={}&\frac{\sqrt6}{F}x.
\label{autotheta}
\end{align}
Here and below a prime denotes $d/dN$. Equations~(\ref{autox})--(\ref{autotheta}), together with Eq.~(\ref{LambdaTheta}), form a regular autonomous system in the variables $(x,y,\Omega_r,\theta)$.



\subsection{Critical points and their cosmological interpretation}
\label{critical_section}

The critical points of the complete system satisfy
\be
x'=y'=\Omega_r'=\theta'=0.
\ee
For finite $F$, Eq.~(\ref{autotheta}) implies $x=0$ at every true equilibrium point. The remaining conditions then separate into the matter- and radiation-dominated critical sets and the extrema of the periodic potential. Because $\theta$ is periodic, we may restrict it to one fundamental interval, $0\leq\theta<2\pi$. The physically relevant critical sets for $0<p<1$ are summarized in Table~\ref{tab:critical}.

\begin{table}[tb]
\centering
\begin{tabular}{|c|c|c|c|c|}
\hline
Critical set & $(x,y,\Omega_r,\theta)$ & $\Omega_m$ & $w_{\rm eff}$ & Interpretation \\
\hline
$M$ & $(0,0,0,\theta_*)$ & $1$ & $0$ & Matter domination \\
\hline
$R$ & $(0,0,1,\theta_*)$ & $0$ & $1/3$ & Radiation domination \\
\hline
$D_{\rm max}$ & $(0,1,0,2k\pi)$ & $0$ & $-1$ & Potential maximum \\
\hline
$D_{\rm min}$ & $(0,1,0,(2k+1)\pi)$ & $0$ & $-1$ & Potential minimum \\
\hline
\end{tabular}
\caption{Critical sets of the autonomous system for $0<p<1$. Here $\theta_*$ is arbitrary for the matter and radiation critical lines, and $k$ is an integer. Since $\theta$ is periodic, one may equivalently take $k=0$ within a fundamental interval.}
\label{tab:critical}
\end{table}

The matter solution $M$ has $\Omega_m=1$ and represents the standard matter-dominated epoch. The radiation solution $R$ has $\Omega_r=1$. Both are one-parameter critical lines because the scalar field is dynamically irrelevant when $x=y=0$, leaving its phase $\theta_*$ undetermined. The dark-energy solutions occur at extrema of the potential, where $\sin\theta_*=0$ and hence $\Lambda_\phi=0$. At both extrema, $x=0$, $y=1$ and $w_{\rm eff}=w_\phi=-1$. Their stability, however, is different because the curvature of the potential changes sign between the maximum and the minimum.

An important consequence of the complete $(x,y,\Omega_r,\theta)$ system is that points with $x\neq0$ cannot be equilibrium points for finite $f_1$, because then $\theta'\neq0$. Thus apparent fixed points obtained by freezing $\Lambda_\phi$ at a nonzero value do not represent fixed points of the full periodic-potential dynamics.

\subsection{Linear stability analysis}
\label{stability_section}

The stability of an isolated hyperbolic critical point is determined by the eigenvalues of the Jacobian matrix of Eqs.~(\ref{autox})--(\ref{autotheta}) \cite{strogatz:2001}. The matter and radiation solutions are critical lines and therefore necessarily possess one zero eigenvalue tangent to the line. Their transverse eigenvalues are sufficient to determine their saddle-type character. At the matter critical line $M$, the eigenvalues are $\left\{0,-\frac32,\frac32,-1\right\}.
\label{eigM} $
The positive and negative transverse eigenvalues show that $M$ is a saddle-type critical set. At the radiation critical line $R$, the eigenvalues are $\left\{0,-1,2,1\right\},
\label{eigR} $
so the radiation solution is also unstable in some directions and acts as a saddle-type early-time critical set.

At the maximum of the potential, $\theta_*=2k\pi$, the eigenvalues are
\be
\mu_1=-3,\qquad \mu_2=-4,
\label{eigmax12}
\ee
and
\be
\mu_{\pm}^{({\rm max})}
=\frac{-3\pm\sqrt{9+\dfrac{12p}{F^2(1+p)}}}{2}.
\label{eigmax}
\ee
For $p>0$, the square root in Eq.~(\ref{eigmax}) is larger than $3$, and consequently one of the two eigenvalues is positive. The de Sitter solution at the potential maximum is therefore a saddle, as expected for a hill-top configuration.

At the minimum, $\theta_*=(2k+1)\pi$, the eigenvalues are
\be
\mu_1=-3,\qquad \mu_2=-4,
\label{eigmin12}
\ee
and
\be
\mu_{\pm}^{({\rm min})}
=\frac{-3\pm\sqrt{9-\dfrac{12p}{F^2(1-p)}}}{2}.
\label{eigmin}
\ee
For $0<p<1$, the real parts of both $\mu_\pm^{({\rm min})}$ are negative. Hence the de Sitter solution at the positive minimum of the axion potential is a future attractor. The result is independent of whether the last pair is real or complex. The stability properties are collected in Table~\ref{tab:stability}.

\begin{table}[tb]
\centering
\begin{tabular}{|c|c|c|}
\hline
Critical set & Eigenvalues & Stability \\
\hline
$M$ & $0,-3/2,3/2,-1$ & Saddle-type critical line \\
\hline
$R$ & $0,-1,2,1$ & Saddle-type critical line \\
\hline
$D_{\rm max}$ & $-3,-4,\mu_+^{({\rm max})},\mu_-^{({\rm max})}$ & Saddle \\
\hline
$D_{\rm min}$ & $-3,-4,\mu_+^{({\rm min})},\mu_-^{({\rm min})}$ & Stable attractor \\
\hline
\end{tabular}
\caption{Linear stability of the critical sets. The quantities $\mu_\pm^{({\rm max})}$ and $\mu_\pm^{({\rm min})}$ are given in Eqs.~(\ref{eigmax}) and (\ref{eigmin}), respectively.}
\label{tab:stability}
\end{table}

For the representative best-fit values quoted in Ref.~\cite{Ruchika:2020avj}, $p=0.57$ and $f_1=0.71$, and in reduced-Planck units where $F=\kappa f_1=f_1$, the maximum has
\be
\left\{-3,-4,0.600,-3.600\right\},
\ee
confirming its saddle nature. At the minimum one finds
\be
\left\{-3,-4,-1.500+2.375i,-1.500-2.375i\right\},
\label{numericmin}
\ee
so the observationally representative late-time solution is a stable focus. Trajectories in its neighbourhood therefore approach the de Sitter state through damped oscillations.

\subsection{Parameter dependence of the late-time attractor}

The parameters $p$ and $f_1$, or equivalently the dimensionless
combination $F=\kappa f_1$, determine the local curvature of the
periodic potential and therefore control the manner in which
trajectories approach the late-time de Sitter attractor. From
Eq.~(\ref{eigmin}), the scalar-sector eigenvalues at the minimum are
given by
\be
\mu_{\pm}^{(\min)}
=
\frac{-3\pm
\sqrt{
9-\dfrac{12p}{F^2(1-p)}
}
}{2}.
\ee
The qualitative nature of the approach is therefore determined by the
discriminant
\be
\Delta=
9-\frac{12p}{F^2(1-p)}.
\label{discmin}
\ee

For
\be
F^2>\frac{4p}{3(1-p)},
\label{nodecondition}
\ee
the discriminant is positive and the two scalar-sector eigenvalues are
real and negative. The de Sitter solution at the minimum is therefore
a stable node, and nearby trajectories approach the attractor
monotonically at the linear level.

On the other hand, when
\be
F^2<\frac{4p}{3(1-p)},
\label{focuscond}
\ee
the discriminant becomes negative and the scalar-sector eigenvalues
form a complex-conjugate pair,
\be
\mu_{\pm}^{(\min)}
=
-\frac{3}{2}\pm i\omega ,
\label{complexeig}
\ee
where
\be
\omega=
\frac{1}{2}
\sqrt{
\frac{12p}{F^2(1-p)}-9
}.
\label{omega}
\ee
The late-time solution is then a stable focus, and trajectories
approach it through damped oscillations. Thus, a larger modulation
amplitude $p$ or a smaller value of $F$ increases the curvature of the
potential around its minimum relative to the Hubble damping and
enhances the oscillatory character of the approach to the attractor.

The boundary between the two regimes is given by
\be
F^2=\frac{4p}{3(1-p)},
\ee
for which $\Delta=0$ and the two eigenvalues coincide at
\be
\mu_{+}^{(\min)}
=
\mu_{-}^{(\min)}
=
-\frac{3}{2}.
\ee
Hence the change across this boundary corresponds to a transition
from a stable node to a stable focus, or equivalently from monotonic
to damped oscillatory relaxation toward the same late-time attractor.

This transition is not a Hopf bifurcation. In the stable-focus regime,
the real part of the complex-conjugate eigenvalue pair remains fixed
at $-3/2$ and therefore never crosses the imaginary axis. Consequently,
no limit cycle is generated. The oscillatory behaviour of the scalar
field and its equation-of-state parameter is instead associated with
damped relaxation around the stable minimum of the periodic potential.

Likewise, no saddle--node bifurcation occurs within the physical
branch $0<p<1$. The extrema of the potential remain located at
\be
\theta=2k\pi
\ee
and
\be
\theta=(2k+1)\pi,
\ee
corresponding respectively to the potential maximum and minimum,
throughout this parameter range. They do not merge or annihilate as
$p$ or $F$ is varied. The maximum remains a saddle, while the positive
minimum remains the stable late-time de Sitter attractor.

In the limit $p\rightarrow1^{-}$, the minimum value of the potential,
\be
V_{\min}=V_0(1-p),
\ee
approaches zero. This marks the boundary of the positive-potential
branch considered here rather than a saddle--node bifurcation.

The resulting phase-space picture is therefore simple. The radiation-
and matter-dominated critical sets describe the transient saddle-type
epochs of the standard cosmological evolution, the de Sitter solution
at the maximum of the potential is unstable, and the positive minimum
provides the stable late-time attractor. Depending on the values of
$(p,F)$, the final approach to this attractor is either monotonic or
damped oscillatory.

\subsection{Numerical phase-space illustrations}
\label{phaseportraits}

To complement the analytical eigenvalue analysis, we numerically integrate the autonomous system and display representative two-dimensional projections of the resulting trajectories. Since the complete system is four dimensional, $(x,y,\Omega_r,\theta)$, these figures should be interpreted as projections of the full cosmological flow rather than as independent two-dimensional autonomous systems. They nevertheless provide a useful visual confirmation of the stability properties derived above.

\begin{figure*}[tb]
\centering
\includegraphics[width=0.48\textwidth]{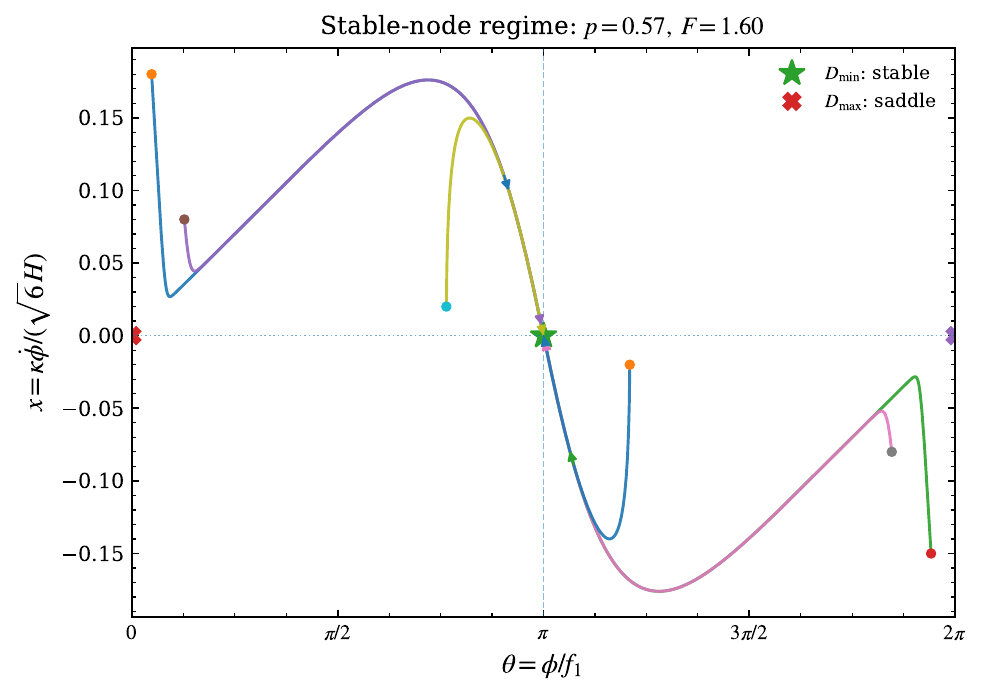}\hfill
\includegraphics[width=0.48\textwidth]{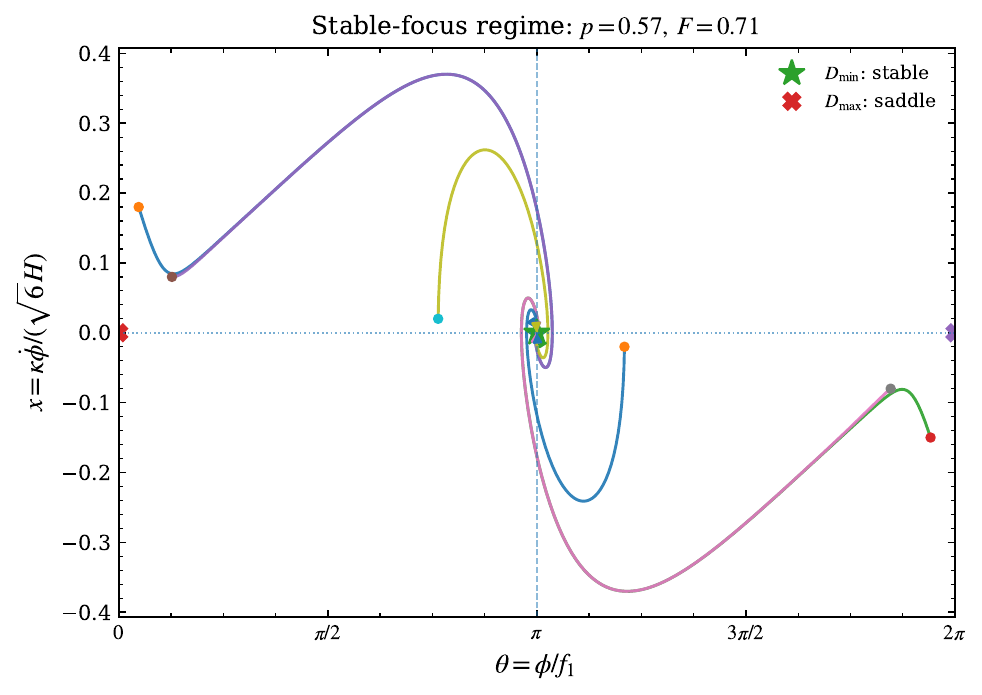}
\caption{{\small Projected trajectories in the $(\theta,x)$ plane for two representative choices of the axion decay-scale parameter. Left: stable-node regime with $p=0.57$ and $F=1.60$. Right: stable-focus regime with $p=0.57$ and $F=0.71$. The point $D_{\rm max}$ at $\theta=0$ modulo $2\pi$ is a saddle, while $D_{\rm min}$ at $\theta=\pi$ modulo $2\pi$ is stable. In the focus regime the trajectories execute damped oscillations around $D_{\rm min}$, whereas in the node regime the approach is non-oscillatory.}}
\label{fig:nodefocus_phase}
\end{figure*}

Figure~\ref{fig:nodefocus_phase} compares the approach to the minimum of the periodic potential in the stable-node and stable-focus regimes. In both panels we use $p=0.57$, while the decay-scale combination $F=\kappa f_1$ is chosen on opposite sides of the boundary in Eq.~(\ref{nodecondition}). The potential maximum, $D_{\rm max}$, occurs at $\theta=0$ modulo $2\pi$ and is a saddle, whereas the minimum, $D_{\rm min}$, occurs at $\theta=\pi$ modulo $2\pi$ and is the late-time attractor. For $F=1.60$ the trajectories approach $D_{\rm min}$ without spiralling, as expected for a stable node. For $F=0.71$ the trajectories wind around the minimum with a decreasing amplitude, providing a direct visualization of the stable-focus behaviour. The latter is precisely the damped oscillatory relaxation predicted by the complex eigenvalue pair in Eq.~(\ref{numericmin}); the trajectories converge toward the fixed point and do not form a closed orbit.

\begin{figure}[tb]
\centering
\includegraphics[width=0.66\columnwidth]{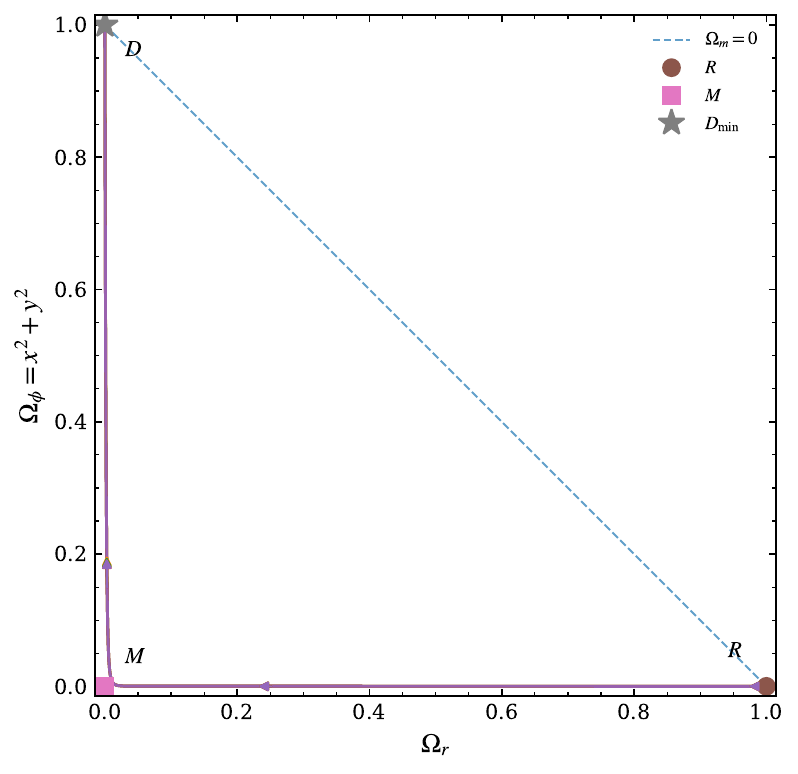}
\caption{{\small Projection of representative cosmological trajectories onto the $(\Omega_r,\Omega_\phi)$ plane. The dashed line denotes the boundary $\Omega_m=0$, or $\Omega_r+\Omega_\phi=1$. The labelled points correspond to radiation domination $R=(1,0)$, matter domination $M=(0,0)$, and the late-time de Sitter attractor $D_{\rm min}=(0,1)$. The trajectories exhibit the standard sequence from radiation domination through a matter-dominated stage to the stable dark-energy solution.}}
\label{fig:omega_phase}
\end{figure}

The global cosmological sequence is illustrated separately in Fig.~\ref{fig:omega_phase} using the $(\Omega_r,\Omega_\phi)$ projection. The physically allowed region satisfies
\be
\Omega_r+\Omega_\phi\leq1,
\ee
where equality corresponds to $\Omega_m=0$. The radiation-dominated state is represented by $R=(1,0)$, the matter-dominated state by $M=(0,0)$, and the final scalar-field-dominated state by $D_{\rm min}=(0,1)$. Representative trajectories leave the vicinity of the radiation point, pass close to the matter-dominated saddle, and subsequently evolve toward the stable dark-energy attractor. The numerical flow therefore reproduces the expected cosmological ordering
\be
R\ \longrightarrow\ M\ \longrightarrow\ D_{\rm min},
\ee
and complements the local fixed-point analysis by demonstrating that the relevant critical sets can be connected by physically admissible cosmological trajectories.

\section{Conclusion}
\label{conclu}

We have studied the dynamical stability of the axion dark-energy model with the periodic potential in Eq.~(\ref{Vphi2}), using the observationally motivated parameter region discussed in Ref.~\cite{Ruchika:2020avj}. The analysis has been formulated directly in terms of the standard normalized variables $x$ and $y$, the radiation density parameter $\Omega_r$, and the periodic field variable $\theta=\phi/f_1$. This choice keeps the physical constraint $\Omega_\phi=x^2+y^2\leq1$ explicit, retains the sign of the scalar-field velocity, and remains regular at $\Omega_\phi=0$.

A central feature of the periodic potential is that its logarithmic slope is field dependent. We therefore retain the exact relation $\Lambda_\phi(\theta)$ and show explicitly that $\Gamma$ is also a function of $\theta$. Evolving the phase variable directly avoids the coordinate difficulty associated with $\Gamma$ at extrema of the potential and ensures that the fixed points are fixed points of the complete scalar-field system rather than of a reduced system with a frozen potential slope.

The radiation- and matter-dominated solutions form saddle-type critical lines and naturally provide transient cosmological epochs. The dark-energy solution located at the maximum of the axion potential is a saddle. By contrast, for the positive-potential branch $0<p<1$, the minimum of the potential is a stable de Sitter attractor with $\Omega_\phi=1$ and $w_\phi=-1$. Thus the model admits the desired qualitative sequence from radiation domination through matter domination to a stable accelerated phase. Numerical phase-space projections confirm this picture: representative trajectories connect the radiation and matter saddle regimes to the final scalar-field attractor, while the $(\theta,x)$ projection clearly distinguishes the non-oscillatory node approach from the damped spiral-like relaxation of the stable-focus regime.

The parameters of the periodic potential determine how the final state is approached. When $F^2>4p/[3(1-p)]$, the attractor is a stable node, whereas for $F^2<4p/[3(1-p)]$ it is a stable focus. The latter case produces damped oscillations of the scalar-field equation of state. Since the real part of the complex eigenvalue pair remains negative and does not cross the imaginary axis, this behaviour is a node-to-focus transition and not a Hopf bifurcation. Similarly, no saddle--node bifurcation occurs among the physical dark-energy extrema in the regular $0<p<1$ phase space.

For the representative values $p=0.57$ and $f_1=0.71$ quoted in Ref.~\cite{Ruchika:2020avj}, the late-time solution lies in the stable-focus regime. The axion field therefore relaxes toward the positive minimum through damped oscillations while preserving a stable asymptotic de Sitter state. These results establish the dynamical consistency of the observationally relevant branch of the model and clarify the origin of its oscillatory late-time behaviour.

\bibliographystyle{unsrt}
\bibliography{sample}

@article{SupernovaSearchTeam:1998fmf,
  author = {Riess, A. G. and others},
  title = {Observational evidence from supernovae for an accelerating universe and a cosmological constant},
  journal = {Astronomical Journal},
  volume = {116},
  pages = {1009--1038},
  year = {1998}
}

@article{SupernovaCosmologyProject:1998vns,
  author = {Perlmutter, S. and others},
  title = {Measurements of Omega and Lambda from 42 high redshift supernovae},
  journal = {Astrophysical Journal},
  volume = {517},
  pages = {565--586},
  year = {1999}
}

@article{Padmanabhan:2002ji,
  author = {Padmanabhan, T.},
  title = {Cosmological constant: The Weight of the vacuum},
  journal = {Physics Reports},
  volume = {380},
  pages = {235--320},
  year = {2003}
}

@article{Johnson:2008se,
  author = {Johnson, M. C. and Kamionkowski, M.},
  title = {Dynamical and Gravitational Instability of Oscillating-Field Dark Energy and Dark Matter},
  journal = {Physical Review D},
  volume = {78},
  pages = {063010},
  year = {2008},
  doi = {10.1103/PhysRevD.78.063010},
  eprint = {0805.1748},
  archivePrefix = {arXiv}
}

@article{Dutta:2008px,
  author = {Dutta, S. and Scherrer, R. J.},
  title = {Evolution of Oscillating Scalar Fields as Dark Energy},
  journal = {Physical Review D},
  volume = {78},
  pages = {083512},
  year = {2008},
  doi = {10.1103/PhysRevD.78.083512},
  eprint = {0805.0763},
  archivePrefix = {arXiv}
}

@article{Joseph:2021omr,
  author = {Joseph, A. and Saha, R.},
  title = {Dark energy with oscillatory tracking potential: observational constraints and perturbative effects},
  journal = {Monthly Notices of the Royal Astronomical Society},
  volume = {511},
  number = {2},
  pages = {1637--1646},
  year = {2022},
  doi = {10.1093/mnras/stac201},
  eprint = {2110.00229},
  archivePrefix = {arXiv}
}

@article{Escamilla:2024fzq,
  author = {Escamilla, L. A. and others},
  title = {Testing an oscillatory behavior of dark energy},
  journal = {Physical Review D},
  volume = {111},
  number = {2},
  pages = {023531},
  year = {2025},
  doi = {10.1103/PhysRevD.111.023531},
  eprint = {2404.00181},
  archivePrefix = {arXiv}
}

@article{Rezaei:2024vtg,
  author = {Rezaei, M.},
  title = {Oscillating Dark Energy in Light of the Latest Observations and Its Impact on the Hubble Tension},
  journal = {Astrophysical Journal},
  volume = {967},
  number = {1},
  pages = {2},
  year = {2024},
  doi = {10.3847/1538-4357/ad3963},
  eprint = {2403.18968},
  archivePrefix = {arXiv}
}

@article{DuttaK:2021,
  author = {Dutta, K. and Maharana, A.},
  journal = {European Physical Journal Special Topics},
  volume = {230},
  pages = {2111},
  year = {2021}
}

@article{Choi:2019jck,
  author = {Choi, G. and Suzuki, M. and Yanagida, T. T.},
  title = {Quintessence axion dark energy and a solution to the hubble tension},
  journal = {Physics Letters B},
  volume = {805},
  pages = {135408},
  year = {2020}
}

@article{Ruchika:2020avj,
  author = {Ruchika and Adil, S. A. and Dutta, K. and Mukherjee, A. and Sen, A. A.},
  title = {Observational constraints on axion(s) dark energy with a cosmological constant},
  journal = {Physics of the Dark Universe},
  volume = {40},
  pages = {101199},
  year = {2023},
  doi = {10.1016/j.dark.2023.101199},
  eprint = {2005.08813},
  archivePrefix = {arXiv}
}

@book{Wainwright:2005,
  author = {Wainwright, J. and Ellis, G. F. R.},
  title = {Dynamical system in Cosmology},
  publisher = {Cambridge University Press},
  year = {2005}
}

@book{Coley:2003,
  author = {Coley, A. A.},
  title = {Dynamical System and Cosmology},
  publisher = {Springer},
  year = {2003}
}

@article{Gunzig:2000,
  author = {Gunzig, E. and others},
  journal = {Classical and Quantum Gravity},
  volume = {17},
  pages = {1783},
  year = {2000}
}

@article{Carot:2003,
  author = {Carot, J. and Collinge, M. M.},
  journal = {Classical and Quantum Gravity},
  volume = {20},
  pages = {707},
  year = {2003}
}

@article{Urena-Lopez:2005pzi,
  author = {Urena-Lopez, L. A.},
  journal = {JCAP},
  volume = {09},
  pages = {013},
  year = {2005}
}

@article{Kumar:2013oda,
  author = {Kumar, S. and Panda, S. and Sen, A. A.},
  journal = {Classical and Quantum Gravity},
  volume = {30},
  pages = {155011},
  year = {2013}
}

@article{Sen:2009yh,
  author = {Sen, S. and Sen, A. A. and Sami, M.},
  journal = {Physics Letters B},
  volume = {686},
  pages = {1--5},
  year = {2010}
}

@article{Fang:2014qga,
  author = {Fang, W. and others},
  journal = {Physical Review D},
  volume = {89},
  pages = {123514},
  year = {2014}
}

@article{Fang:2014bta,
  author = {Fang, W. and others},
  journal = {European Physical Journal C},
  volume = {76},
  pages = {492},
  year = {2016}
}

@article{Roy:2014yta,
  author = {Roy, N. and Banerjee, N.},
  journal = {European Physical Journal Plus},
  volume = {129},
  pages = {162},
  year = {2014}
}

@article{Roy:2014wqa,
  author = {Roy, N. and Banerjee, N.},
  eprint = {1412.0837},
  archivePrefix = {arXiv},
  year = {2014}
}

@article{Mishra:2019uqm,
  author = {Mishra, S. and Chakraborty, S.},
  journal = {Modern Physics Letters A},
  volume = {34},
  year = {2019}
}

@article{Das:2023rat,
  author = {Das, S. and others},
  journal = {Physical Review D},
  volume = {108},
  pages = {083517},
  year = {2023}
}

@article{Kamionkowski:2014zda,
  author = {Kamionkowski, M. and Pradler, J. and Walker, D. G. E.},
  journal = {Physical Review Letters},
  volume = {113},
  pages = {251302},
  year = {2014}
}

@article{Dutta:2008qn,
  author = {Dutta, S. and Scherrer, R. J.},
  journal = {Physical Review D},
  volume = {78},
  pages = {123525},
  year = {2008}
}

@book{strogatz:2001,
  author = {Strogatz, S. H.},
  title = {Nonlinear Dynamics and Chaos},
  publisher = {Westview Press},
  year = {2001}
}

@article{Thanankullaphong:2026anl,
    author = "Thanankullaphong, Phusuda and Sahoo, Prasanta and Hassan Puttasiddappa, Prajwal and Roy, Nandan",
    title = "{Quintom Dark Energy: Future Attractor and Phantom Crossing in Light of DESI DR2 Observation}",
    eprint = "2601.02284",
    archivePrefix = "arXiv",
    primaryClass = "astro-ph.CO",
    month = "1",
    year = "2026"
}

@article{Roy:2026jcm,
  author = {Roy, N. and Urena-Lopez, L. A.},
  title = {Dynamical Systems in Cosmology: Reviewing An Alternative Approach},
  year = {2026},
  eprint = {2603.24568},
  archivePrefix = {arXiv}
}

@article{Planck2020,
    author = "Aghanim, N. and others",
    collaboration = "Planck",
    title = "{Planck 2018 results. VI. Cosmological parameters}",
    eprint = "1807.06209",
    archivePrefix = "arXiv",
    primaryClass = "astro-ph.CO",
    doi = "10.1051/0004-6361/201833910",
    journal = "Astron. Astrophys.",
    volume = "641",
    pages = "A6",
    year = "2020",
    note = "[Erratum: Astron.Astrophys. 652, C4 (2021)]"
}

@article{BAO2017,
   title={The clustering of galaxies in the completed SDSS-III Baryon Oscillation Spectroscopic Survey: cosmological analysis of the DR12 galaxy sample},
   volume={470},
   ISSN={1365-2966},
   url={http://dx.doi.org/10.1093/mnras/stx721},
   DOI={10.1093/mnras/stx721},
   number={3},
   journal={Monthly Notices of the Royal Astronomical Society},
   publisher={Oxford University Press (OUP)},
   author={Alam, Shadab and Ata, Metin and Bailey, Stephen and Beutler, Florian and Bizyaev, Dmitry and Blazek, Jonathan A. and Bolton, Adam S. and Brownstein, Joel R. and Burden, Angela and Chuang, Chia-Hsun and et al.},
   year={2017},
   month={Mar},
   pages={2617–2652}
}

@article{BBN2021,
   title={Completed SDSS-IV extended Baryon Oscillation Spectroscopic Survey: Cosmological implications from two decades of spectroscopic surveys at the Apache Point Observatory},
   volume={103},
   ISSN={2470-0029},
   url={http://dx.doi.org/10.1103/PhysRevD.103.083533},
   DOI={10.1103/physrevd.103.083533},
   number={8},
   journal={Physical Review D},
   publisher={American Physical Society (APS)},
   author={Alam, Shadab and Aubert, Marie and Avila, Santiago and Balland, Christophe and Bautista, Julian E. and Bershady, Matthew A. and Bizyaev, Dmitry and Blanton, Michael R. and Bolton, Adam S. and Bovy, Jo and et al.},
   year={2021},
   month={Apr}
}

@article{DES2018,
    author = "Abbott, T. M. C. and others",
    collaboration = "DES",
    title = "{Dark Energy Survey year 1 results: Cosmological constraints from galaxy clustering and weak lensing}",
    eprint = "1708.01530",
    archivePrefix = "arXiv",
    primaryClass = "astro-ph.CO",
    reportNumber = "FERMILAB-PUB-17-294-PPD",
    doi = "10.1103/PhysRevD.98.043526",
    journal = "Phys. Rev. D",
    volume = "98",
    number = "4",
    pages = "043526",
    year = "2018"
}

@article{Sh0ES2019,
   title={Large Magellanic Cloud Cepheid Standards Provide a 1\% Foundation for the Determination of the Hubble Constant and Stronger Evidence for Physics beyond $\Lambda$CDM},
   volume={876},
   ISSN={1538-4357},
   url={http://dx.doi.org/10.3847/1538-4357/ab1422},
   DOI={10.3847/1538-4357/ab1422},
   number={1},
   journal={The Astrophysical Journal},
   publisher={American Astronomical Society},
   author={Riess, Adam G. and Casertano, Stefano and Yuan, Wenlong and Macri, Lucas M. and Scolnic, Dan},
   year={2019},
   month={May},
   pages={85}
}

@article{Riess_2022,
	doi = {10.3847/1538-4357/ac8f24},
  
	url = {https://doi.org/10.3847%2F1538-4357%2Fac8f24},
  
	year = 2022,
	month = {oct},
  
	publisher = {American Astronomical Society},
  
	volume = {938},
  
	number = {1},
  
	pages = {36},
  
	author = {Adam G. Riess and Louise Breuval and Wenlong Yuan and Stefano Casertano and Lucas M. Macri and J. Bradley Bowers and Dan Scolnic and Tristan Cantat-Gaudin and Richard I. Anderson and Mauricio Cruz Reyes},
  
	title = {Cluster Cepheids with High Precision Gaia Parallaxes, Low Zero-point Uncertainties, and Hubble Space Telescope Photometry},
  
	journal = {The Astrophysical Journal}
}

@article{Wong:2019kwg,
    author = "Wong, Kenneth C. and others",
    title = "{H0LiCOW \textendash{} XIII. A 2.4 percent measurement of H0 from lensed quasars: 5.3\ensuremath{\sigma} tension between early- and late-Universe probes}",
    eprint = "1907.04869",
    archivePrefix = "arXiv",
    primaryClass = "astro-ph.CO",
    doi = "10.1093/mnras/stz3094",
    journal = "Mon. Not. Roy. Astron. Soc.",
    volume = "498",
    number = "1",
    pages = "1420--1439",
    year = "2020"
}

@article{Roy:2024kni,
    author = "Roy, Nandan",
    title = "{Dynamical dark energy in the light of DESI 2024 data}",
    eprint = "2406.00634",
    archivePrefix = "arXiv",
    primaryClass = "astro-ph.CO",
    month = "6",
    year = "2024"
}

@article{DESI:2024aqx,
    author = "Calderon, R. and others",
    collaboration = "DESI",
    title = "{DESI 2024: Reconstructing Dark Energy using Crossing Statistics with DESI DR1 BAO data}",
    eprint = "2405.04216",
    archivePrefix = "arXiv",
    primaryClass = "astro-ph.CO",
    month = "5",
    year = "2024"
}

@article{Roy:2025cxk,
    author = "Roy, Nandan and Chakrabarti, Soumya",
    title = "{Is Phantom Barrier Crossing Inevitable? A Cosmographic Analysis}",
    eprint = "2508.13740",
    archivePrefix = "arXiv",
    primaryClass = "astro-ph.CO",
    month = "8",
    year = "2025"
}

@article{Cortes:2024lgw,
    author = "Cort{\^e}s, Marina and Liddle, Andrew R.",
    title = "{Interpreting DESI's evidence for evolving dark energy}",
    eprint = "2404.08056",
    archivePrefix = "arXiv",
    primaryClass = "astro-ph.CO",
    doi = "10.1088/1475-7516/2024/12/007",
    journal = "JCAP",
    volume = "12",
    pages = "007",
    year = "2024"
}

@article{shlivko2024assessing,
  title={Assessing observational constraints on dark energy},
  author={Shlivko, David and Steinhardt, Paul J},
  journal={Physics Letters B},
  volume={855},
  pages={138826},
  year={2024},
  publisher={Elsevier}
}

@article{Ishak:2024jhs,
   title={Modified gravity constraints from the full shape modeling of clustering measurements from DESI 2024},
   volume={2025},
   ISSN={1475-7516},
   url={http://dx.doi.org/10.1088/1475-7516/2025/09/053},
   DOI={10.1088/1475-7516/2025/09/053},
   number={09},
   journal={Journal of Cosmology and Astroparticle Physics},
   publisher={IOP Publishing},
   author={Ishak, M and Pan, J and Calderon, R and Lodha, K and Valogiannis, G and Aviles, A and Niz, G and Yi, L and Zheng, C and Garcia-Quintero, C and others},
   year={2025},
   month=sep, pages={053} }

@article{Filbert:2023zhg,
    author = "Filbert, S. and others",
    title = "{Broad absorption line quasars in the Dark Energy Spectroscopic Instrument Early Data Release}",
    eprint = "2309.03434",
    archivePrefix = "arXiv",
    primaryClass = "astro-ph.CO",
    doi = "10.1093/mnras/stae1610",
    journal = "Mon. Not. Roy. Astron. Soc.",
    volume = "532",
    number = "4",
    pages = "3669--3681",
    year = "2024"
}

@article{DESI:2024uvr,
    author = "Adame, A. G. and others",
    collaboration = "DESI",
    title = "{DESI 2024 III: Baryon Acoustic Oscillations from Galaxies and Quasars}",
    eprint = "2404.03000",
    archivePrefix = "arXiv",
    primaryClass = "astro-ph.CO",
    reportNumber = "FERMILAB-PUB-24-0159-PPD",
    month = "4",
    year = "2024"
}

@article{DESI:2025zgx,
    author = "Abdul Karim, M. and others",
    collaboration = "DESI",
    title = "{DESI DR2 Results II: Measurements of Baryon Acoustic Oscillations and Cosmological Constraints}",
    eprint = "2503.14738",
    archivePrefix = "arXiv",
    primaryClass = "astro-ph.CO",
    reportNumber = "FERMILAB-PUB-25-0169-PPD",
    month = "3",
    year = "2025"
}

@article{Wright:2025xka,
    author = "Wright, Angus H. and others",
    title = "{KiDS-Legacy: Cosmological constraints from cosmic shear with the complete Kilo-Degree Survey}",
    eprint = "2503.19441",
    archivePrefix = "arXiv",
    primaryClass = "astro-ph.CO",
    doi = "10.1051/0004-6361/202554908",
    journal = "Astron. Astrophys.",
    volume = "703",
    pages = "A158",
    year = "2025"
}

@article{DESI:2025fii,
    author = "Lodha, K. and others",
    collaboration = "DESI",
    title = "{Extended dark energy analysis using DESI DR2 BAO measurements}",
    eprint = "2503.14743",
    archivePrefix = "arXiv",
    primaryClass = "astro-ph.CO",
    reportNumber = "FERMILAB-PUB-25-0164-PPD",
    doi = "10.1103/w4c6-1r5j",
    journal = "Phys. Rev. D",
    volume = "112",
    number = "8",
    pages = "083511",
    year = "2025"
}

@article{DESI:2024hhd,
    author = "Adame, A. G. and others",
    collaboration = "DESI",
    title = "{DESI 2024 VII: cosmological constraints from the full-shape modeling of clustering measurements}",
    eprint = "2411.12022",
    archivePrefix = "arXiv",
    primaryClass = "astro-ph.CO",
    reportNumber = "FERMILAB-PUB-24-0854-PPD",
    doi = "10.1088/1475-7516/2025/07/028",
    journal = "JCAP",
    volume = "07",
    pages = "028",
    year = "2025"
}

@article{Shlivko:2024llw,
    author = "Shlivko, David and Steinhardt, Paul J.",
    title = "{Assessing observational constraints on dark energy}",
    eprint = "2405.03933",
    archivePrefix = "arXiv",
    primaryClass = "astro-ph.CO",
    doi = "10.1016/j.physletb.2024.138826",
    journal = "Phys. Lett. B",
    volume = "855",
    pages = "138826",
    year = "2024"
}

@article{Wolf:2023uno,
    author = "Wolf, William J. and Ferreira, Pedro G.",
    title = "{Underdetermination of dark energy}",
    eprint = "2310.07482",
    archivePrefix = "arXiv",
    primaryClass = "astro-ph.CO",
    doi = "10.1103/PhysRevD.108.103519",
    journal = "Phys. Rev. D",
    volume = "108",
    number = "10",
    pages = "103519",
    year = "2023"
}

@article{Wolf:2024eph,
    author = "Wolf, William J. and Garc{\'\i}a-Garc{\'\i}a, Carlos and Bartlett, Deaglan J. and Ferreira, Pedro G.",
    title = "{Scant evidence for thawing quintessence}",
    eprint = "2408.17318",
    archivePrefix = "arXiv",
    primaryClass = "astro-ph.CO",
    doi = "10.1103/PhysRevD.110.083528",
    journal = "Phys. Rev. D",
    volume = "110",
    number = "8",
    pages = "083528",
    year = "2024"
}

@article{Yuan:2022jqf,
    author = "Yuan, Sihan and Garrison, Lehman H. and Eisenstein, Daniel J. and Wechsler, Risa H.",
    title = "{Stringent {\ensuremath{\sigma}}8 constraints from small-scale galaxy clustering using a hybrid MCMC~+~emulator framework}",
    eprint = "2203.11963",
    archivePrefix = "arXiv",
    primaryClass = "astro-ph.CO",
    doi = "10.1093/mnras/stac1830",
    journal = "Mon. Not. Roy. Astron. Soc.",
    volume = "515",
    number = "1",
    pages = "871--896",
    year = "2022"
}

@article{Simon:2022lde,
    author = "Simon, Th{\'e}o and Zhang, Pierre and Poulin, Vivian and Smith, Tristan L.",
    title = "{Consistency of effective field theory analyses of the BOSS power spectrum}",
    eprint = "2208.05929",
    archivePrefix = "arXiv",
    primaryClass = "astro-ph.CO",
    doi = "10.1103/PhysRevD.107.123530",
    journal = "Phys. Rev. D",
    volume = "107",
    number = "12",
    pages = "123530",
    year = "2023"
}

@article{DES:2021vln,
    author = "Secco, L. F. and others",
    collaboration = "DES",
    title = "{Dark Energy Survey Year 3 results: Cosmology from cosmic shear and robustness to modeling uncertainty}",
    eprint = "2105.13544",
    archivePrefix = "arXiv",
    primaryClass = "astro-ph.CO",
    reportNumber = "FERMILAB-PUB-21-253-AE, DES-2019-0480",
    doi = "10.1103/PhysRevD.105.023515",
    journal = "Phys. Rev. D",
    volume = "105",
    number = "2",
    pages = "023515",
    year = "2022"
}

@article{Busch:2022pcx,
    author = "Busch, J. L. van den and others",
    title = "{KiDS-1000: Cosmic shear with enhanced redshift calibration}",
    eprint = "2204.02396",
    archivePrefix = "arXiv",
    primaryClass = "astro-ph.CO",
    doi = "10.1051/0004-6361/202142083",
    journal = "Astron. Astrophys.",
    volume = "664",
    pages = "A170",
    year = "2022"
}

@article{Planck:2015lwi,
    author = "Ade, P. A. R. and others",
    collaboration = "Planck",
    title = "{Planck 2015 results. XXIV. Cosmology from Sunyaev-Zeldovich cluster counts}",
    eprint = "1502.01597",
    archivePrefix = "arXiv",
    primaryClass = "astro-ph.CO",
    doi = "10.1051/0004-6361/201525833",
    journal = "Astron. Astrophys.",
    volume = "594",
    pages = "A24",
    year = "2016"
}
\end{document}